# Data Hiding in Speech Signal Using Steganography and Encryption

Hanisha Chowdary N
School of Electronics and Engineering
Vellore Institute of Technology
Vellore, India
hanishachowdaryn@gmail.com

Karan K
School of Electronics and Engineering
Vellore Institute of Technology
Vellore, India
kaushikkaran2904@gmail.com

Bharath K P
School of Electronics and Engineering
Vellore Institute of Technology
Vellore, India
bharathkp25@gmail.com

Rajesh Kumar M *Senior Member, IEEE*
School of Electronics and Engineering
Vellore Institute of Technology
Vellore, India
mrajeshkumar@vit.ac.in

*Abstract*— Data privacy and data security are always on highest priority in the world. We need a reliable method to encrypt the data so that it reaches the destination safely. Encryption is a simple yet effective way to protect our data while transmitting it to a destination. The proposed method has state of art technology of steganography and encryption. This paper puts forward a different approach for data hiding in speech signals. A ten-digit number within speech signal using audio steganography and encrypting it with a unique key for better security. At the receiver end the same unique key is used to decrypt the received signal and then hidden numbers are extracted. The proposed approach performance can be evaluated by PSNR, MSE, SSIM and bit-error rate. The simulation results give better performance compared to existing approach.

## I. INTRODUCTION

Communication is an imperative need of individuals, for projection of ideas, to interact, for connectivity across the world, the purpose narrows down to enhancing the quality of life. The Speech is most dominant and essential method of communication between people. Innumerable techniques are utilized in speech processing for diverse purposes for example, speech synthesis, speech recognition, vocal dialog, speech transmission, speech enhancement, and speech coding. Amongst the linearly growing curve of technology and networking, information is more susceptible to threats, such as modifying the information in the channel or intercepted by unauthorized. Steganography is widely used in different applications as image processing, signal processing, speech processing and multimedia applications. The encryption enables for data transfer where the data must only be accessed by authorized personal. It is in the scope of paper to provide an approach which delivers the speech signal in such a way that it would be known to receiver upon the information being modified, and in case of the signal being intercepted it would be encrypted there by allowing only authorized personal access to the hidden information.

There have been many methods followed to conceal a particular data in a speech signal. [1] Adopted to hide the data in the silence intervals of the speech. This is done by altering the number of samples in silence intervals. Audio watermarking was used to prevent the direct and indirect attacks on speaker recognition system from the unauthorized user [2]. [3] K. Vimal and S. A. kather discussed about extraction of data hidden in the silence region of speech signal by non-voiced detection algorithm. The methodology proposed by Subir and Dr. Amit M. Joshi [3] in their paper uses an innovative approach of Discrete Cosine Transform (DCT) and Discrete Wavelet Transform (DWT) for the purpose of watermarking and they incorporated Arnold transformation and error correction technique [4] to increase functioning of the approach. Images are extensively used for process of hiding data in steganography, as audio steganography is perplexing because of higher precision in Human Auditory system (HAS) as compared to Human Visual System(HVS) [5].

Temporal and transform domain techniques are frequently exploited for audio steganography. Temporal domain contains LSB encoding, parity coding and echo hiding, whereas transform domain is associated with frequency and wavelet domain techniques. [6][7] Provides an inference that techniques in wavelet domain have higher hiding capacity and transparency. LSB method is employed in our algorithm. [8] Explains different audio steganography algorithms, which were developed to be the source of information hidden for different communication technologies. [9] It gives the summary of watermarking techniques on the basis of robustness and imperceptibility. [10] Explains the data hiding in the mid frequency of an image and generated the randomize key for encryption. This paper incorporates the method of steganography and encryption in speech signal. The input speech signal is divided into N number of frames. In the LSB of each frame information is hidden. The unique key which is known to authorize users is used for encryption of the stenographic signal. Then the proposed approach is compared with existing method.

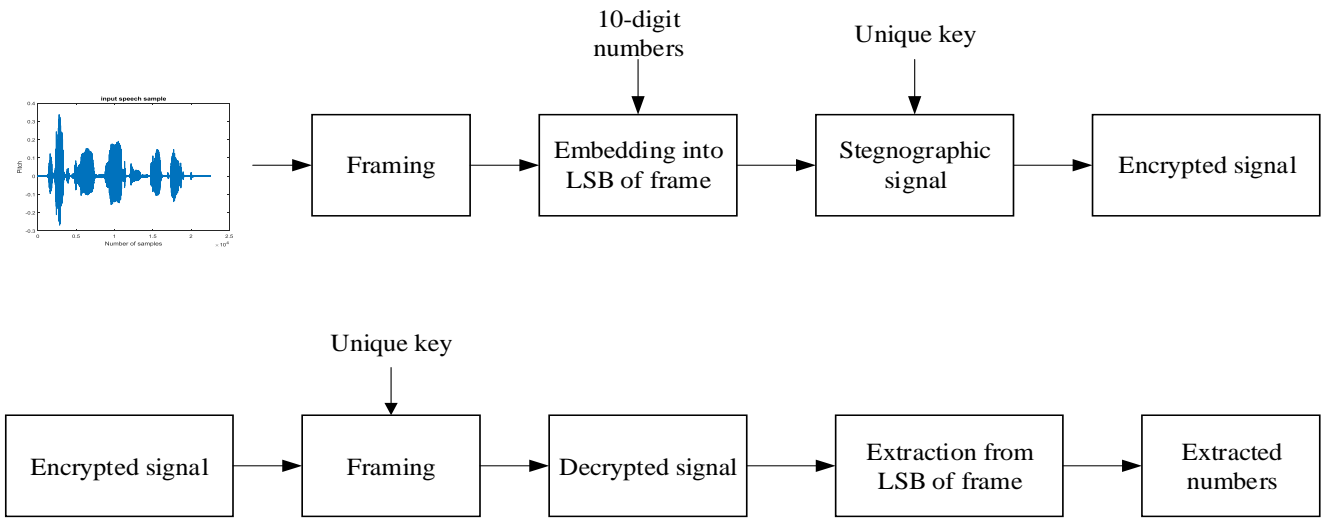

Fig.1 Proposed block diagram of Stegnographic and encrypted signal

The rest of the paper is organized as follows: Section-II explains the proposed methodology to encrypt the ten-digit number into the signal and secret key. Simulation results are discussed and compared with existing approach in section III. Section IV provides the conclusion.

## II. PROPSED MEthodology

In our proposed method, the speech signal is sampled at 8 kHz with the frame length of 160 samples. In the LSB of each frame information is hidden. The unique key which is known to authorized users is used for encryption of the stenographic signal. It empowers for a more secure communication, in the case of external unauthorized tampering, there will be a variance between the sequence numbers sent and received. Signal is protected from spoofing due to the scrambling algorithm utilized. The proposed block diagram is as shown in the fig. 1.

A. Proposed procedure for encryption

Step 1: **Framing:** It is a process of dividing speech signal into N number of frames with the length of 160 samples per frame. Frames of the speech signal are shown in the fig.2.

Step 2: **Embedding**: A set of N integers with length of 10 digits which are varying from 0 to 9. The generated numbers are divided by 1000 to ensure that the amplitude of the signal after later step of encryption remains in the desired range. The obtained floating-point numbers are concealed in the 10 samples of LSB of each frame. The number of LSB samples can be chosen according to the bits in need of hiding, here 10 is chosen as we are embedding a 10-digit number. Frames of the speech signal after steganography are shown in fig.3.

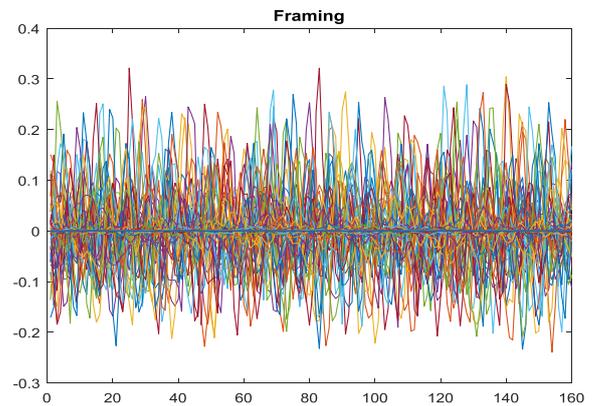

Fig.2 Frames of speech signal.

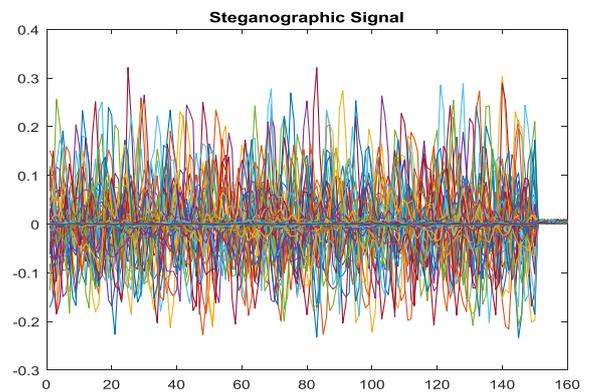

Fig.3 Frames of Stegnographic signal

Step 3: **Encryption:** Foremost aspect of the encryption lies in generation of unique key, several logics can be applied, an insight into cryptography would guide in development of key with compound logic ensuring it's not cracked in case of interception. The frames of speech signal after encryption are shown in fig.4.

Below, is the pseudo code for generation of key, which is used in this paper. The generated key of size 160 bits, is applied to each frame of the stenographic signal for a improved security in communication. The data format of the signal with key embedded would be decimal.

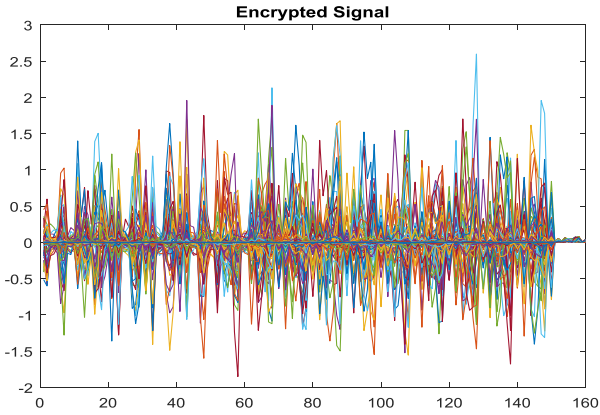

Fig.4 Frames of Encrypted Signal

**Process for key generation:**
Choose any 3 prime numbers p, q and m
k = remainder (p*q, 10)
p_new = (k+1)*m
q_new = p/q + positional value of the current sample
p = p_new
q= q_new
Repeat for N number of frames

If the value of k becomes zero, then k is incremented by 1. If the p/q gives a floating value it must be converted into closest integer.

B. Proposed procedure for decryption

Step 1: **Framing:** Received encrypted signal divided into N number of frames of equal size similar to the process in encryption, as framing is the necessary stage in processing a signal.

Step 2: **Decryption:** the framed signal is decrypted by applying the unique key generated previously in encryption process which is known to authorized users alone. Frames of the decrypted speech signal are shown in fig.5.

Step 3: **Extraction phase:** the hidden sequence of numbers is extracted from LSB samples of each frame. The extracted numbers are compared with the initially generated numbers, the transmission is declared successful upon matching, and failure in other case. In the case of failure, there arises a suspicion of attack on the signal; it can be further investigated by checking the instances of mismatch, and other factors to eliminate the cause of noise, which currently is beyond the scope of this paper. In the case of failure, there arises a suspicion of attack on the signal, it can be further investigated by checking the instances of mismatch, and other factors to eliminate the cause of noise, which currently is beyond the scope of this paper.

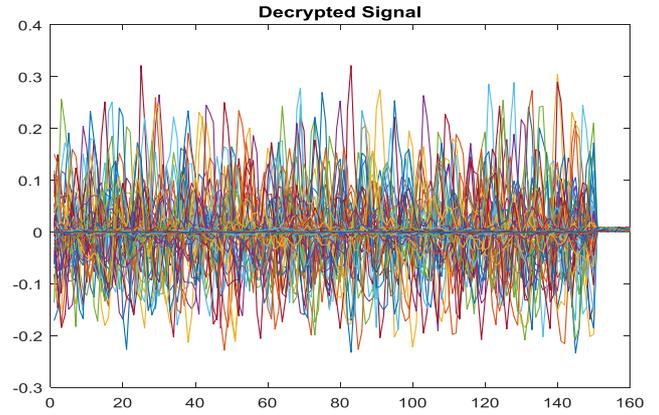

Fig.5 Frames of Decrypted signal

III. SIMULATION RESULTS

We simulated the proposed approach in MATLAB 2016a. The analysis of the proposed approach is assessed by Peak Signal to Noise Ratio (PSNR), Mean Square Error (MSE), SSIM and Bit error rate. These metrics are used to calculate the robustness of the proposed method. Fig. 6 to show the input speech sample, fig.7 shows the stenographic signal and fig.8 has the extracted signal. Bit error rate is the ratio of number of error bits to the total number of bits transmitted [11]. It is expressed as:

$$Bit\ error\ rate\ (\%) = \frac{error\ bits}{total\ number\ of\ bits} * 100 \qquad (1)$$

For accurate results of above parameters, 20 different audio speech signals were chosen from the database of speech signals, obtained from AURORA with the duration between 2.5 to 5 sec, chosen signals are of clean speech. Two different methods are applied to each signal, the method of steganography in samples of silence interval followed by the encryption of the signal and the method of steganography in samples of all frames followed by encryption of signal. The values of MSE, PSNR, SSIM and BER are calculated for each signal after each of the above 2 methods, and the averaged values of each parameter are considered.

Fig.6 show the input clean speech sample which chosen form the above mentioned database which is nearly having 18000

samples. Fig. 7 shows the stegnographic signal where data is hidden in the LSBs of each frame. Fig.8 shows the encrypted speech signal, a unique is applied to encrypt the stegnographic signal for better secure communication. The length of key size is 160 bits.

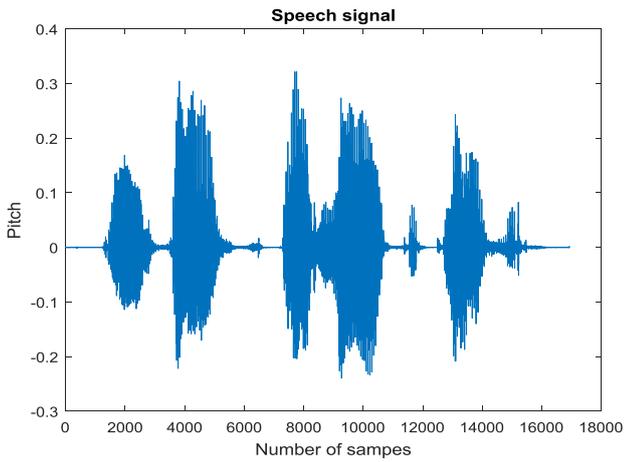

Fig.6: Input speech signal

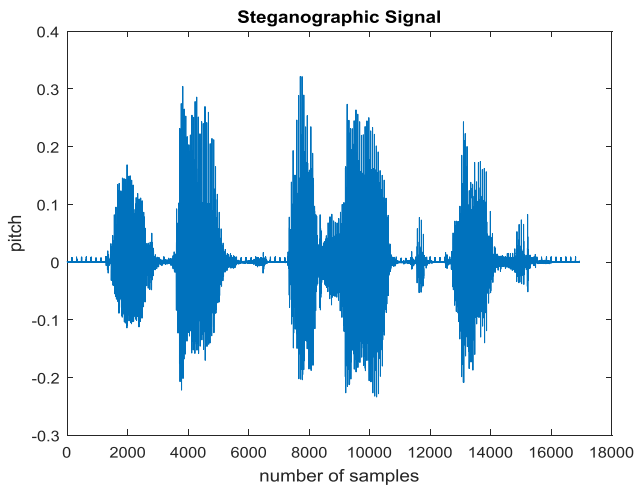

Fig.7: steganographic signal

The comparison results of proposed and existing method are tabulated in Table 1, values of MSE, PSNR and SSIM of 20 samples are tabulated. Table 2 contains value of BER of proposed and silence interval method. From the values tabulated, it is evident that proposed approach is giving lower MSE by $0.74 \times 10^{-4}$, PSNR and SSIM elevated by 2.3864 and 0.0296 respectively than obtained from method of silence interval. The percentage of bit error rate obtained in our approach is zero, so this approach is useful where the communication targeted is in demand of better values of parameters mentioned above and would require transmission of signal that is neither intercepted nor modified in the channel before reaching desired receiver.

The percentage of bit error rate obtained in our approach is zero, so this approach is useful where the communication targeted is in demand of better values of parameters mentioned above and would require transmission of signal that is neither intercepted nor modified in the channel before reaching desired receiver.

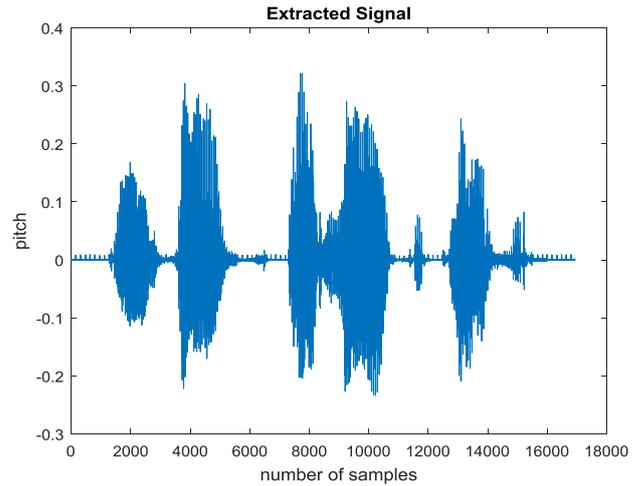

Fig.8: Extracted speech signal

Table 1: Comparison of steganography in silence intervals with proposed method

| Methods | MSE | PSNR | SSIM |
|---|---|---|---|
| Silence interval [1] | $1.83 \times 10^{-4}$ | 37.2965 | 0.9375 |
| Proposed | $1.09 \times 10^{-4}$ | 39.6829 | 0.9671 |

Table 2: Comparison of Bit Error Rate (BER) of data hiding in silence method with proposed approach

| Method | BER(%) |
|---|---|
| Silence interval [1] | 10.414 |
| Proposed Method | 0.0 |

## IV. CONCLUSION

This paper presents a novel method involving steganography and encryption thereby ensuring a relatively reliable communication without spoofing. The data is embedded in the LSB samples of frames. The unique key is applied to the stenographic signal for better secure transmission. The effect on the quality of the retrieved signal is minimal and there is zero percent of error in the bits extracted. Advancement of the current work can be done by hiding alphanumeric characters.

In the future work different types of data can be hidden at the LSBs of each frame and also different types of attacks or noises can be applied to check the accuracy of the algorithm.

REFERENCES

[1] Mohammad Shirali-Shahreza and Sajad Shirali Shahreza, "Steganography in Silence Intervals of Speech", Institute of Electrical and Electronics Engineers (IEEE), International Conference on Intelligent Information Hiding and Multimedia Signal Processing, 2008.

[2] Nihalkumar Desai and Nikunj Tahilramani, "Digital Speech Watermarking for Authenticity of Speaker in Speaker Recognition System", (2016) International Conference on Micro-Electronics and Telecommunication engineering

[3] K. Vimal and S. A. kather, " Real Steganography in Non Voice Part of the Speech", International Journal of Computer Applications, Volume 46, May, 2012

[4] Subir and Dr. Amit M. Joshi, "DWT-DCT based Blind Audio Watermarking using Arnold Scrambling and Cyclic Codes", (2016) 3rd International Conference on Signal Processing and Integrated Networks (SPIN)

[5] Abdul, W., Carré, P., & Gaborit, P. (2013). "Error correcting codes for robust color wavelet watermarking." EURASIP Journal on Information Security, 2013(1), 1-17.

[6] N. Cvejic, "Algorithms for Audio Watermarking and Steganography", Oulu University Press, Finland, 2004

[7] N. Cvejic and T. Seppanen, "A wavelet domain LSB insertion algorithm for high capacity audio steganography," *Proceedings of 10th IEEE Digital Signal Processing Workshop and 2nd Signal Processing Education Workshop*, October 2002, pp. 53- 55.

[8] Fatiha Djebbar, Beghdad Ayad Karim, Abed Meraim and Habib Hamam, "Comparative Study of Digital Audio Steganography Techniques," EURASIP journal on audio, speech and music processing, springer, 2012.

[9] S. A. R. Al-Haddad and Nematollahi, Mohammad Ali, "An overview of digital speech watermarking." International Journal of Speech Technology, 2013, pp. 1-18.

[10] Bharath k. p and Shubhratha S, "SDEM Based ATM Card Number Hiding Using R-G-B Randomize Key Generator", IRJET, 2016.

[11] Papa Kostas, et al, Moment based local image watermarking via genetic optimization, Applies Mathematics and Computation, 2014